\documentclass[journal]{IEEEtran}

\usepackage{amsmath,amssymb}
\usepackage{cite}
\usepackage{siunitx}
\usepackage{booktabs}
\usepackage[hidelinks]{hyperref}
\usepackage{xurl}            % URLを任意位置で安全に改行
\hypersetup{breaklinks=true} % 2段組でも改行を許可

\usepackage{graphicx}% Include figure files
\usepackage{dcolumn}% Align table columns on decimal point
\usepackage{bm}% bold math
\usepackage{color}% color text
\usepackage{setspace} 
\usepackage{comment}
\usepackage{physics}
\usepackage{ulem}
\usepackage{braket}
\usepackage[linesnumbered, ruled, vlined]{algorithm2e}

\newcommand{\mr}[1]{\mathrm{#1}}
\DeclareMathOperator{\supp}{supp}

\begin{document}

\title{Efficient implementation of randomized quantum algorithms with dynamic circuits}

\author{
  \IEEEauthorblockN{%
    Shu~Kanno\IEEEauthorrefmark{1}\IEEEauthorrefmark{2},\quad
    Ikko~Hamamura\IEEEauthorrefmark{3},\quad
    Rudy~Raymond\IEEEauthorrefmark{2}\IEEEauthorrefmark{3},\quad
    Qi~Gao\IEEEauthorrefmark{1}\IEEEauthorrefmark{2},\quad
    and~Naoki~Yamamoto\IEEEauthorrefmark{2}\IEEEauthorrefmark{4}%
  }
  \\
  \IEEEauthorblockA{\IEEEauthorrefmark{1}
    Mitsubishi Chemical Corporation, Science \& Innovation Center,\\
    1000 Kamoshida-cho, Aoba-ku, Yokohama 227-8502, Japan}
  \\
  \IEEEauthorblockA{\IEEEauthorrefmark{2}
    Quantum Computing Center, Keio University,\\
    Hiyoshi 3-14-1, Kohoku, Yokohama 223-8522, Japan}
  \\
  \IEEEauthorblockA{\IEEEauthorrefmark{3}
    IBM Quantum, IBM Research – Tokyo,\\
    19-21 Nihonbashi Hakozaki-cho, Chuo-ku, Tokyo 103-8510, Japan}
  \\
  \IEEEauthorblockA{\IEEEauthorrefmark{4}
    Department of Applied Physics and Physico-Informatics, Keio University,\\
    3-14-1 Hiyoshi, Kohoku-ku, Yokohama, Kanagawa 223-8522, Japan}
  \\
\thanks{S.K. and I.H. contributed equally to this work. 
    The present address of I.H. is NVIDIA G.K., 2-7-3 Akasaka, Minato-ku, Tokyo 107-0052, Japan.
    R.R. is an advisor to WPI Bio2Q, Keio University, Tokyo 160-8582, Japan. 
    Corresponding author: Shu Kanno (email: shu.kanno@quantum.keio.ac.jp). 
    A part of this work was performed for Council for Science, Technology and Innovation (CSTI), Cross-ministerial Strategic Innovation Promotion Program (SIP), “Promoting the application of advanced quantum technology platforms to social issues”(Funding agency: QST). This work was supported by the MEXT Quantum Leap Flagship Program under Grants No. JPMXS0118067285 and No. JPMXS0120319794.}
}

\maketitle

\textbf{\begin{abstract}
Randomized algorithms are crucial subroutines in quantum computing, but the requirement to execute many types of circuits on a real quantum device has been challenging to their extensive implementation. 
In this study, we propose an engineering method to reduce the executing time for randomized algorithms using dynamic circuits, i.e., quantum circuits involving intermediate measurement and feedback processes. 
The main idea is to generate the probability distribution defining a target randomized algorithm on a quantum computer, instead of a classical computer, which enables us to implement a variety of static circuits on a single dynamic circuit with many measurements. 
We applied the proposed method to the task of random Pauli measurement for one qubit on an IBM superconducting device, showing that a 14,000-fold acceleration of executing time was observed compared with a conventional method using static circuits. 
Additionally, for the problem of estimating expectation values of 28- and 40-qubit hydrogen chain models, we successfully applied the proposed method to realize the classical shadow with 10 million random circuits, which is the largest demonstration of classical shadow.
This work significantly simplifies the execution of randomized algorithms on real quantum hardware.
\end{abstract}}


%\IEEEpeerreviewmaketitle

\section{\label{sec:introduction}Introduction}

\IEEEPARstart{R}{andomized} algorithms are important subroutines in quantum computing. Their applications span a wide range of areas, including expectation value calculations~\cite{Huang2020-ld,Hadfield2020-cr, Huang2021-uf,Wu2023-zq, Gresch2025-jd}, error mitigation~\cite{Cai2022-sc}, time evolution~\cite{Campbell2019-sx,Yang2021-ie}, and circuit cutting~\cite{Peng2020-fl,Harada2024-iz}. 
Here, we take the classical shadow~\cite{Huang2020-ld} as an example. 
The classical shadow is a technique that calculates expectation values of $M$ (local) observables using $\order{\log M}$ measurements through random measurements. 
For instance, in quantum chemistry calculations, the number of Pauli terms in a Hamiltonian scales as $\order{N_\mr{so}^4}$, and thus the classical shadow can suppress the polynomial growth in the number of measurements due to $\order{\log{N_\mr{so}}}$, where $N_\mr{so}$ is the number of spin orbitals.

\begin{figure*}[]
    \centering
    \includegraphics[width=1\textwidth]{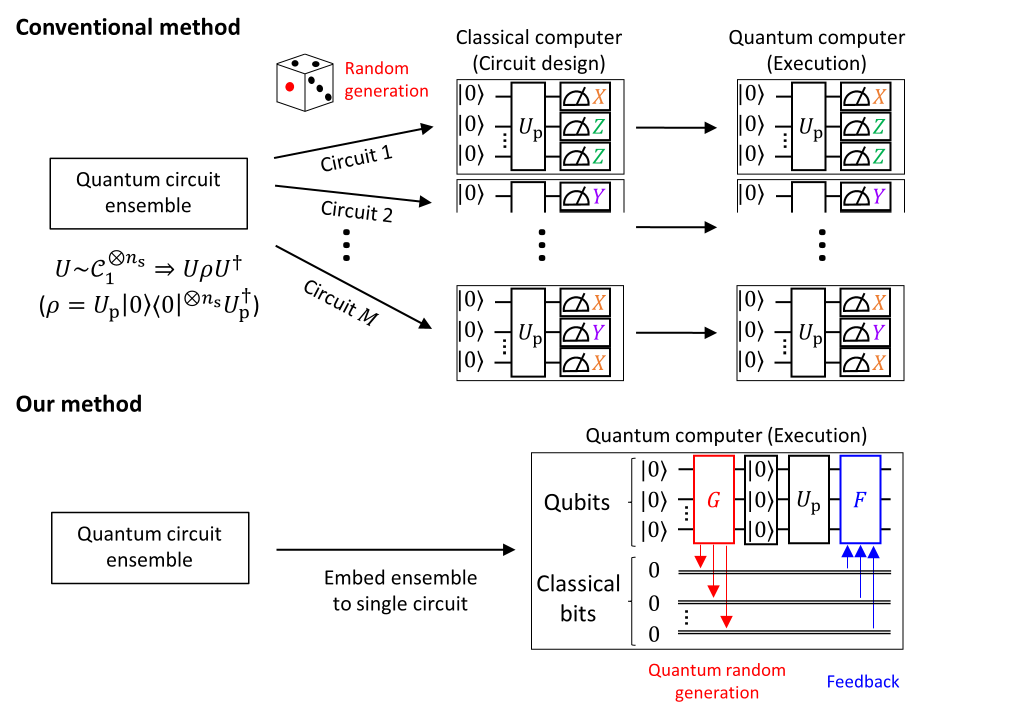}
    \caption{Overview of our method for the case of random Pauli measurement. 
    $\mathcal{C}_1$ is a single-qubit Clifford group, and $n_\mr{s}$ is the number of qubits. 
    Single lines and double lines in the quantum circuits represent quantum and classical registers (bits), respectively. 
    The conventional methods randomly generate multiple static circuits and execute each of them with one (or a few) shot(s) on a quantum computer. 
    In contrast, the proposed method executes a single dynamic circuit multiple times. }    
    \label{FIG1_Overview_2column.pdf}
\end{figure*}

However, execution time is a major obstacle to validating randomized algorithms on real devices. 
This is because randomized algorithms require the execution of a huge variety of circuits generated from a probability distribution. 
For example, in the classical shadow using single-qubit Clifford operations, qubits in a circuit are randomly measured in the $X$, $Y$, or $Z$ bases according to the uniform distribution for each shot. 
However, under a fixed number of circuit executions, performing many shots on a single circuit is faster than executing one shot for each of multiple circuits, due to the compilation process of each circuit~\cite{He2023-at}, where in this research, ``compilation'' refers to pulse compilation, i.e., the conversion from native gate sequence to pulse instruction schedule for the device, and does not include transpilation as circuit optimization and gate decomposition.
Concretely, the basic execution capacity on an IBM device had been limited to 300 circuits per job, whereas 100,000 shots per circuit are allowed~\cite{Korhonen2024-cp} (while recently the limitation seems to have been alleviated~\cite{UnknownUnknown-sy}). 
Consequently, executing a large number of circuits requires splitting them across multiple jobs. 
Even experiments with a few qubits would require about $10^4$ circuits~\cite{Levy2024-de, Zhang2021-zf, Struchalin2021-bc}, and each job can take tens of minutes to hours~\cite{Korhonen2024-cp}, preventing such randomized algorithm from becoming practical. 

In other words, the obstacle arises because the compilation time before executing circuits becomes long. Moreover, additional delays occur if the quantum circuits are not executed until the compilation of all circuits packed into a single job has been completed (as the current IBM specification~\cite{UnknownUnknown-vd}). Note that the circuit stitching, executing multiple circuits using qubit reset and restart~\cite{UnknownUnknown-tr}, may reduce the runtime, but it still requires compilation for each of the circuits, so the obstacle still remains. Therefore, the reduction of the compilation time is crucial to validate randomized algorithms on real devices.

To address this engineering challenge, we propose an accelerated implementation method of randomized quantum algorithms. Our idea is based on the dynamic circuit~\cite{Baumer2024-ak, Carrera-Vazquez2024-yd}, which is a protocol involving mid-circuit measurements with feedback, commonly used in applications such as error correction. 
An overview of our approach is shown in Fig.~\ref{FIG1_Overview_2column.pdf}, where we depict the case of random Pauli measurement.
In conventional methods shown in the upper panel, multiple circuits to be executed are first designed classically according to the probability distribution corresponding to a target randomized algorithm followed by circuit compilations, and they are each executed with one (or a few) shot(s) on a quantum computer. 
In contrast, our proposed method shown in the lower panel uses only a single dynamic circuit. 
This circuit first generates the probability distribution in the quantum register state and temporarily stores the measurement results in classical registers ($G$ in the figure); the stored results are fed back to the quantum registers ($F$) for the re-initialized state ($U_{\mr{p}}$ with qubit reset). 
The values stored in the classical register vary for each shot, enabling the execution of different static circuits for each shot on a single dynamic circuit. 
Since the circuit submitted to the device is a single dynamic circuit, the compilation is only one time, which allows for rapid sampling of a large number of types of static circuits exceeding the circuit limit that can be packed into a single job. 
Therefore, the proposed method would enable a drastic reduction of the total execution time for randomized algorithms.
Note that our method is applicable because dynamic circuits are supported on other types of devices~\cite{Mayer2024-yo,Caune2024-gx, Google-Quantum-AI-and-Collaborators2025-hl}. Superconducting type devices may benefit even more than particle trap devices due to their fast gate operation~\cite{Nielsen2010-hw}.
We also mention that our implementation has a lower requirement for the time between measurement and feedback than error correction.

The rest of this paper is composed as follows. 
The technical details of our method are explained in Sec.~\ref{sec:methods}. Verification for the single-qubit random Pauli measurement is shown in Sec.~\ref{sec:single qubit verification}. 
Energy calculations for chemical models by using the classical shadow are demonstrated in Sec.~\ref{sec:Large scale demonstrations}. 
Conclusions and outlooks are stated in Sec.~\ref{sec: conclusions}. 

\begin{figure*}[]
    \centering
    \includegraphics[width=0.6\textwidth]{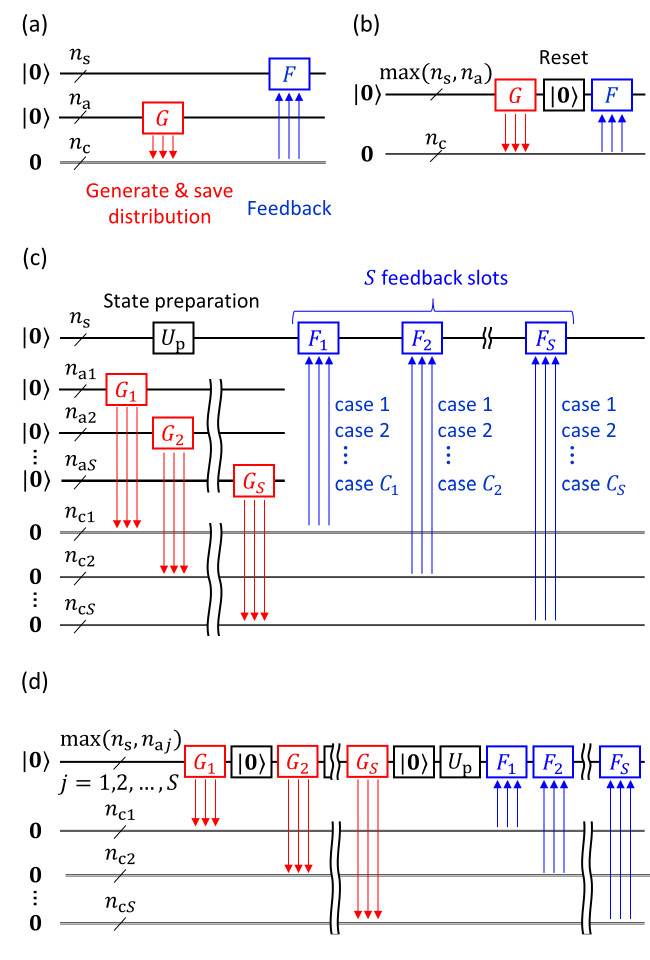}
    \caption{Depictions of the proposed method. 
    Single lines and double lines represent quantum and classical registers, respectively. 
    (a) Abstract circuit diagram. (b) Qubit saving version. (c) Detailed circuit diagram. 
    In (a), the first single line represents system qubits and the second single line represents ancilla qubits, while the system and ancilla qubits are shared in (b). (d) Qubit saving version of (c).}
    \label{FIG2_Method_less1.5column.pdf}
\end{figure*}

\section{\label{sec:methods}Methods}
A high-level sketch of a circuit in our method is shown in Fig.~\ref{FIG2_Method_less1.5column.pdf}(a), where the circuit is composed of $n_\mr{s}$ system qubits, $n_\mr{a}$ ancilla qubits, and $n_\mr{c}$ classical bits. 
We first encode the probability distribution corresponding to a target randomized algorithm, to a quantum state; we measure them and record the results on classical bits ($G$, red in the figure).
Then, we execute conditional feedback operations based on the measured results ($F$, blue in the figure). 
Given a fixed number of circuit executions, executing an identical circuit with multiple shots is faster than executing many different circuits with one (or a few) shot, and thus our method is expected to run randomized algorithms with much less time compared to the conventional methods. 
Note that to obtain the results of $N$ shots for one circuit corresponding to a given random number outcome, after storing a bitstring in a classical register, one can repeat the execution and reset of the system circuit with feedback using the stored bitstring $N$ times.

While the circuit in Fig.~\ref{FIG2_Method_less1.5column.pdf}(b) requires $n_\mr{s} + n_\mr{a}$ qubits, the qubit overhead can be suppressed by reusing ancilla qubits using the qubit reset in Fig.~\ref{FIG2_Method_less1.5column.pdf}(b). The qubit count becomes $\max(n_\mr{s}, n_\mr{a})$, where $\abs{n_\mr{s} - n_\mr{a}}$ qubits are not used during an ancilla or a system part.
In the next section, we adopt this circuit, because it can reduce not only the number of qubits but also possible noise affections on the system qubits during the idle time (i.e., the period of executing on $G$) in the case of Fig.~\ref{FIG2_Method_less1.5column.pdf}(a).
Note that there are no explicit order relation among $n_{\mr{s}}$, $n_{\mr{a}}$, and $n_{\mr{c}}$ in general.

Let us consider a detail of the circuit Fig.~\ref{FIG2_Method_less1.5column.pdf}(a), which is shown as Fig.~\ref{FIG2_Method_less1.5column.pdf}(c).
The circuit includes $S$ feedback slots $\{F_1, \ldots, F_S\}$  and each slot $F_j$ randomly executes one of the $C_j$ gate operations depending on the measurement result on $G_j$.
Corresponding to each $F_{j}$, we assume to use $n_{\mr{a}j}$ qubits for generating the probability distribution that is the output of $G_{j}$, where $n_\mr{a}=\sum_{j=1}^S n_{\mr{a}j}$. 
The distribution can be encoded either analytically or numerically, for example, using a variational procedure~\cite{Nakaji2022-xe} for the latter case. 
The measurement result is stored in the $n_{\mr{c}j}$ classical bit, where $n_\mr{c}=\sum_{j=1}^S n_{\mr{c}j}$. 
Note that $n_\mr{a}$ and $n_\mr{c}$ are at most $S \lceil \mr{max}_j(\log_2 (C_j)) \rceil$.
Here, we depicted a state preparation gate $U_{\mr{p}}$ explicitly to make it easier to understand while it can be generally included in $F_1$. 

Fig.~\ref{FIG2_Method_less1.5column.pdf}(d) presents a qubit saving version of the Fig.~\ref{FIG2_Method_less1.5column.pdf}(c) and the circuit details of the figure (b). 
While the relationship between $n_{\mr{a}j}$ and $n_\mr{s}$ depends on the size of the support of the classical probability distribution, certain algorithms of interest can take $n_\mr{s} \geq n_{\mr{a}j}$, in which case there is no qubit overhead. Examples of such algorithms, as described in detail at the end of this section, include the single-qubit random Clifford, qDrift~\cite{Campbell2019-sx}, gate cutting, and error mitigation.
Additionally, the coherent time (i.e., executable gate count) of the system is expected to remain unaffected by the probability generation since the operations on the system qubits are executed after the qubit reset.
Therefore, unless there are system-level constraints on quantum circuit execution time, it would be efficient to use the schemes in Figs.~\ref{FIG2_Method_less1.5column.pdf}(b) and (d) rather than general schemes in the figures (a) and (c) for these algorithms.

\begin{figure*}[]
    \centering
    \includegraphics[width=1.5\columnwidth]{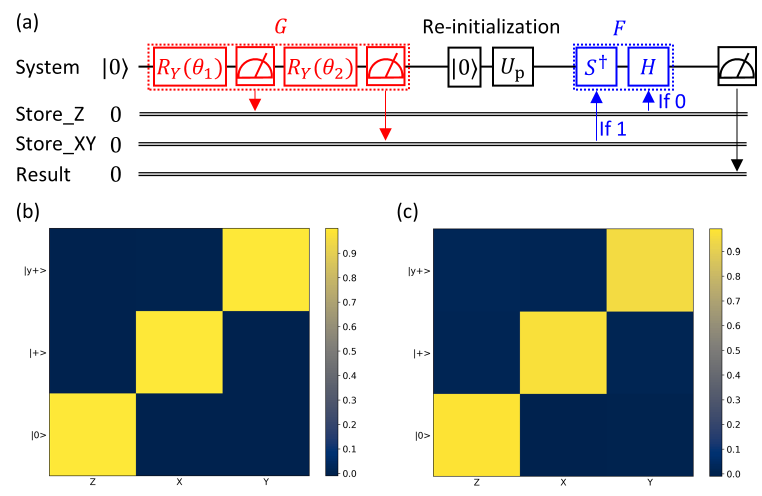}
    \caption{One qubit test for the random Pauli measurement. (a) Circuit for the verification. The single line and double lines represent quantum and classical registers, respectively. The rectangle areas surrounded by red and blue dotted lines represent $G$ and $F$, respectively. (b) and (c) show the contour plots for the expectation values of $Z$, $X$, and $Y$ for respective initialized states $\ket{0}, \ket{+}$, and $\ket{y+}$. (b) State vector simulator results. (c) Real device results ($ibmq\_kolkata$).}
    \label{FIG3_One_qubit_test_1column.pdf}
\end{figure*}

We can efficiently execute these circuits when $S$ is $\order{\mr{poly}(n_\mr{s})}$ since the circuit depth becomes exponential when $S = \Omega(e^{n_\mr{s}})$. Regarding $C_j$, if one naively generates $C_j = 2^{n_{\mr{c}j}}$ cases from the $n_{\mr{c}j}$ classical bits, $C_j$ is restricted to be $\order{\mr{poly}(n_\mr{s})}$. However, if a sophisticated implementation is possible, such as the bitwise case branching shown in Fig.~\ref{FIG3_One_qubit_test_1column.pdf}(a) of the next section, the circuit can be implemented even when $C_j$ is $O(e^{n_\mr{s}})$.

One executable example is qDrift a random compiler for Hamiltonian simulation suitable for chemical models. If the target Hamiltonian $H$ is of the form $H=\sum_{i=1}^L h_i H_i$ with $h_i$ a positive real coefficient and $H_i$ a Hermitian operator (e.g., Pauli string), then $S$ and $C_j$ scale as $(\sum_{i=1}^L h_i)^2 = \order{\mr{poly}(n_\mr{s})}$ and $L = \order{\mr{poly}(n_\mr{s})}$, respectively. This can be executed by no-qubit overhead in the naive implementation, and see Appendix~\ref{sec: construction example of qDRIFT} for details.
In contrast, the naive implementation of the $n_\mr{s}$-qubit random Clifford version of the classical shadows algorithm requires $C_j = \order{2^{\mr{poly}(n_\mr{s})}}$ with $S = 1$. Thus, the possibility of a no qubit overhead ($n_\mr{s} \geq n_{\mr{a}j}$) implementation remains open and will likely require a sophisticated procedure for choice of the random operation~\cite{Koenig2014-kr, van-den-Berg2020-aq}. Yet, as demonstrated in the next section, the single-qubit random Clifford version of the classical shadows admits a no-qubit-overhead implementation using $S=n_\mr{s}$ and $C_j = 3$ for Pauli $X, Y$ and $Z$ basis measurements. Other implementable examples are the gate cutting and error mitigation: a two-qubit gate is decomposed into a sum of 16 terms of single-qubit gate channels, which can be implemented with $n_{\mr{a}j} = \log_2 16 = 4$, and the error mitigation procedure is almost the same as the gate cutting~\cite{Endo2018-nt}.

\begin{figure*}[]
    \centering
    \includegraphics[width=1\textwidth]{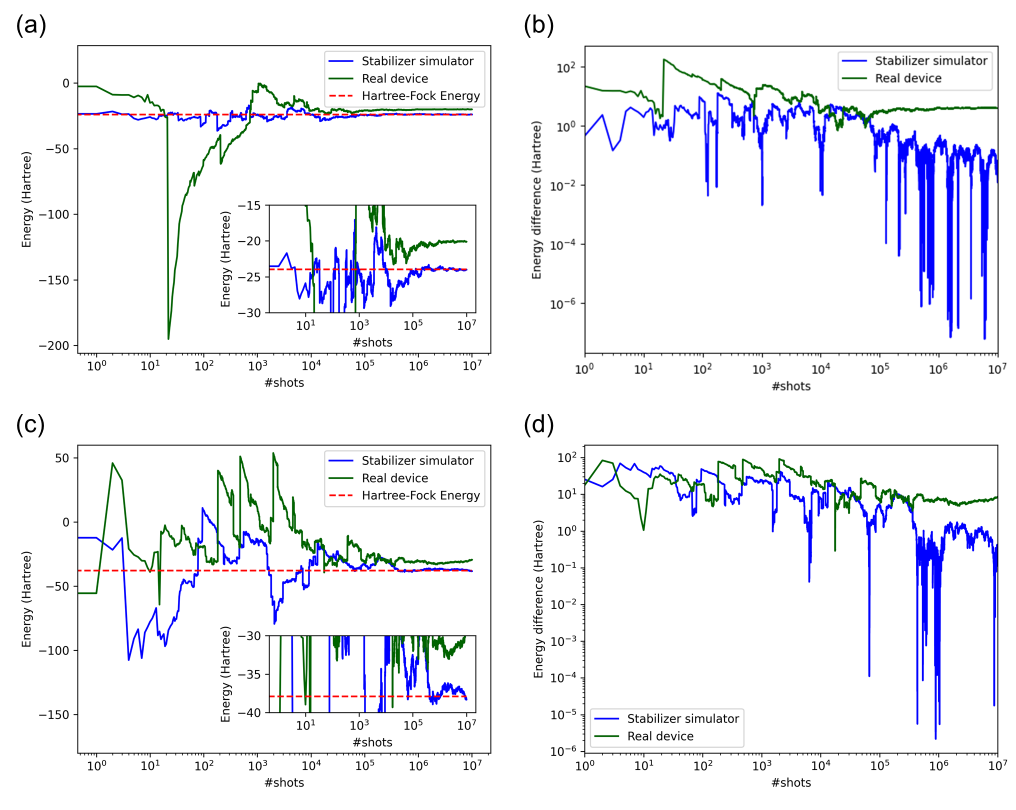}
    \caption{
    Results of expectation values for the hydrogen chain model with Hartree--Fock state. 
    (a) and (b) [(c) and (d)] are the results for 28 [40] qubits. 
    (a) and (c) show the exact values of energy (Hartree-Fock energy) $\bra{{\rm HF}}H\ket{{\rm HF}}$ as well as the estimated values via the real devices and the simulator. 
    For the real device experiment, the readout error mitigation was not carried out. The insets are enlarged views of the y-axis.
    (b) and (d) show the results of differences between the exact values and the estimated values in the log-log plot. 
    We adopt $ibm\_kawasaki$ and $ibm\_kyiv$ devices for the cases of 28 and 40 qubits demonstration, respectively. 
    }
    \label{FIG4_Large_system_result_2column.pdf}
\end{figure*}

\section{\label{sec:Results and Discussions} Results and Discussions}

\subsection{\label{sec:single qubit verification}
Verification with random measurement for single qubit}

First, we demonstrate our method with the random measurements in the $X$, $Y$, and $Z$ bases for a single qubit.
Figure~\ref{FIG3_One_qubit_test_1column.pdf}(a) shows the constructed circuit. 
This circuit consists of one quantum and three classical registers. 
The system and ancilla qubits are shared as in Fig.~\ref{FIG2_Method_less1.5column.pdf}(a) ($n_\mr{s}=n_\mr{a}=1$). 
The first two classical resisters store the probability distribution ($n_\mr{c}=2$), and the last classical resister stores a result to calculate expectation values.

While we can consider the implementation with $S=1$ and $C_1=3$ like Fig.~\ref{FIG2_Method_less1.5column.pdf}(c), we here implement the circuit that involves fewer feedback operations, as follows (see the naive implementation in Appendix~\ref{sec: straightforward implementation for random Pauli measurement}). 
The RY gate operation followed by measurement are repeated twice at the beginning, with the first- and second-measurement results stored in the classical registers ``$\mr{Store\_Z}$'' and ``$\mr{Store\_XY}$'', respectively (red in Fig.~\ref{FIG3_One_qubit_test_1column.pdf}(a)).

More precisely, we set the rotation angles of the two RY gates to $\theta_1 = 2\arccos\sqrt{2/3}$ and $\theta_2 = 2\arccos\sqrt{1/2}$, which generates $|\phi_1\rangle$ followed by $|\phi_2\rangle$: 
\begin{equation}
\label{single qubit example}
    |\phi_1\rangle = \sqrt{\frac{2}{3}}|0\rangle + \sqrt{\frac{1}{3}}|1\rangle, ~~ 
    |\phi_2\rangle = \pm\sqrt{\frac{1}{2}}|0\rangle + \sqrt{\frac{1}{2}}|1\rangle,
\end{equation}
where plus (minus) sign appears in $\ket{\phi_2}$ when $\ket{0}$ ($\ket{1}$) was measured in the first measurement of $\ket{\phi_1}$.
%}
The probabilities of the measurement result are given by 
\begin{eqnarray*}
  && {\rm Pr}(\mr{Store\_Z}=0, \mr{Store\_XY}=0)=\frac{2}{3}\cdot\frac{1}{2}=\frac{1}{3}, \\
  && {\rm Pr}(\mr{Store\_Z}=0, \mr{Store\_XY}=1)=\frac{2}{3}\cdot\frac{1}{2}=\frac{1}{3}, \\ && {\rm Pr}(\mr{Store\_Z}=1, \mr{Store\_XY}=0)=\frac{1}{3}\cdot\frac{1}{2}=\frac{1}{6}, \\ && {\rm Pr}(\mr{Store\_Z}=1, \mr{Store\_XY}=1)=\frac{1}{3}\cdot\frac{1}{2}=\frac{1}{6}.
\end{eqnarray*}

Next, the state is reset, and an initial state $|\psi\rangle$ is generated by $U_{\mr{p}}$. 
Then, corresponding to the above measurement results from top to bottom, we apply $H, HS^\dagger, S^\dagger$, and $I$ to $|\psi\rangle$, respectively (blue in Fig.~\ref{FIG3_One_qubit_test_1column.pdf}(a)). 
Finally, we measure $Z$, meaning that we make the measurement of $X, Y, Z$, and $Z$, respectively, corresponding to the above (pre)measurement result, because of $HZH=X$, $SHZHS^{\dag}=Y$, $SZS^{\dag}=Z$, and $IZI=Z$. 
As a consequence, 
\begin{eqnarray*}
  && {\rm Pr}({\rm measure}~X)={\rm Pr}(\mr{Store\_Z}=0, \mr{Store\_XY}=0)=\frac{1}{3}, \\
  && {\rm Pr}({\rm measure}~Y)={\rm Pr}(\mr{Store\_Z}=0, \mr{Store\_XY}=1)=\frac{1}{3}, \\ 
  && {\rm Pr}({\rm measure}~Z)={\rm Pr}(\mr{Store\_Z}=1) =\frac{1}{3},
\end{eqnarray*}
thus we realize the random measurement of $X, Y, Z$ with equal probability $1/3$. 

Finally, the measured result is stored in the classical bit ``$\mr{Result}$''.
We can estimate the expected value for the randomized measurement by using the selected basis stored in $\mr{Store\_Z}$ and $\mr{Store\_XY}$ in addition to the measured result in $\mr{Result}$.
Note that if the ratio of measurement bases needs to be changed, which happens in the Locally-Biased Classical Shadow~\cite{Hadfield2020-cr} for example, it can be adjusted by modifying the RY rotation angles.

We calculate the expectation values $\expval{Z}$, $\expval{X}$, and $\expval{Y}$ based on their respective eigenstates $\ket{0}$, $\ket{+}$, and $\ket{y+}$ using $U_{\mr{p}}=I,H$, and $R_X(-\pi/2)$. 
The whole quantum computations were conducted based on Qiskit~\cite{Javadi-Abhari2024-qm}.
The results obtained using the QASM simulator and the real device $ibmq\_kolkata$ are presented in Figs.~\ref{FIG3_One_qubit_test_1column.pdf}(b) and (c), respectively. 
The values in each row were obtained by executing a single type of dynamic circuit for 100,000 shots.
In both the figures, only the expectation values corresponding to the eigenstates are almost 1, while the others are almost 0.
The execution time for these 100,000 quantum circuits was 40 seconds, which is obtained from the value of time\_taken in a result output. 
For comparison with the conventional approach, we executed 100 random measurement circuits by 1 shot for 100 jobs; then the execution took 540 seconds. 
Therefore, the proposed method achieved a speedup of $\frac{100000}{40} / \frac{100}{540} \approx 14,000$ times. Note that as in Appendix~\ref{sec: execution time analysis}, we found the slowdown in the conventional approach is likely due to the compilation time rather than the pure execution time of quantum circuits~\cite{Karalekas2020-kd, Dalvi2024-bq}.

We discuss the noise overhead in our implementations. A possible source of the overhead introduced by ancilla qubits is the error arising from deviations of the measured probability distribution from the ideal one; for the example in this section, if the measurement error rate is $e$, the probabilities ${\rm Pr}({\rm measure}~X), {\rm Pr}({\rm measure}~Y)$, and ${\rm Pr}({\rm measure}~Z)$ change from the ideal $1/3$ to $(1-e/2)/3, (1-e/2)/3$ and $(1+e)/3$, respectively. However, we can check the measurement outcomes from the ancilla qubits for each shot. Thus, for algorithms with few outcomes, specifically, $n_{\mr{c}j} = \order{\log n_{\mr{s}}}$, this discrepancy can be witnessed and rectified by rejection sampling, which gives an actual shot count reduction, although the rejection sampling would be impractical for the distribution with $\Omega(\mr{poly}(n_\mr{s}))$ outcomes. We mention that this discrepancy can arise due to errors introduced by the quantum implementation of the classical sampling procedure, including measurement errors and errors that scale with the complexity of the classical sampling process.

We finally comment on the potential for further accelerations. In our current setup, each job contained only a single circuit. Allowing the execution of multiple circuits per job could lead to additional speedups. However, IBM’s imposed instruction count limit~\cite{UnknownUnknown-sy} —-- which heavily penalizes certain operations, such as measurement and reset --- poses challenges for scaling this approach. As a result, as shown in later sections, we were forced to limit each job to one circuit and reduce the number of shots in some cases due to the limitation.
Nevertheless, this limitation may be temporary. Dynamic circuits were introduced primarily to support error correction, suggesting that these restrictions could be relaxed in the future. Furthermore, recently introduced batch jobs, i.e., parallel compilation as described in the following section offer another avenue for accelerating computation~\cite{UnknownUnknown-ql}.

\subsection{\label{sec:Large scale demonstrations}Large scale demonstrations}

Here we demonstrate the effectiveness of our method on relatively large-scale chemical problems. 
Specifically, we calculated the expectation values of Hamiltonians of hydrogen models, using the classical shadow with random Pauli measurement. 
We study the model of (14e, 14o) and (20e, 20o), corresponding to 28 and 40 qubits, respectively, where ($A$e, $B$o) represents $A$ electrons and $B$ orbitals in a chemical model.
The 28 (40) qubit hydrogen chain model was created by arranging 14 (20) hydrogen atoms at one angstrom interval. The STO-3G basis set was used. 
The constructed second-quantized Hamiltonian $H$ is represented as
\begin{equation}
    H = \sum_{ij} c_{ij} a_i^\dagger a_j + \sum_{ijkl} c_{ijkl} a_i^\dagger a_j^\dagger a_l a_k,
\end{equation}
which is transformed into a qubit Hamiltonian as
\begin{equation}
    H = \sum_{I} C_{I} P_I,
\end{equation}
where $c_{ij}$, $c_{ijkl}$, and $C_{I}$ are coefficients, $a_i$ ($a_i^\dagger$) is creation (annihilation) operator, and $P_{I}$ is a Pauli string.
The second-quantized Hamiltonian was constructed from the result of the Hartree--Fock calculation, which was transformed by the Bravyi--Kitaev transformation to obtain the qubit Hamiltonian. 
The state in the re-initialization is the Hartree--Fock state; $\ket{\rm HF}=U_{\rm p}\ket{0}^{\otimes n_{\mr{s}}} = \prod_{i \in \mr{occupied}} X_i \ket{0}^{\otimes n_{\mr{s}}}$.

To estimate the expectation value $\langle H \rangle=\sum_{I} C_{I} \langle P_I \rangle$, in the Pauli-based classical shadow formalism $\langle P_I \rangle$ is calculated as the mean of ${\rm Tr}(P_I \hat{\rho})$ with $\hat{\rho}$ a random state called the snapshot: 
\[
     \hat{\rho}=\bigotimes_{i=1}^n \left[ 3U_i^\dagger |b_i\rangle \langle b_i|U_i - I \right]. 
\]
$U_i$ randomly takes the single-qubit Pauli operator $X, Y$, or $Z$ with equal probability $1/3$, and $\bigotimes_i|b_i\rangle$ is the projected computational basis state obtained as a result of measuring $\bigotimes_i U_i |HF\rangle$. 
We construct the quantum circuit together with the classical register as a straightforward extension of that studied in the previous subsection; i.e., product of the circuit shown in Fig.~\ref{FIG3_One_qubit_test_1column.pdf}(a).
The detail of the circuit configuration for the 28 and 40 qubits are shown in Appendix~\ref{sec:large circuits}. 
We used the Eagle devices, specifically $\mathit{ibm\_kawasaki}$ and $\mathit{ibm\_kyiv}$ for 28 and 40 qubit models, where the numbers of shots (i.e., the numbers of snapshots) per job are 100,000 and 50,000, respectively. 
Again, we emphasize that only one circuit was packed for each job.
A single job took around 30 minutes while we partially used the batch jobs to reduce the total runtime. The optimization level for a circuit was three. 
We also classically performed the state preparation and the random Pauli measurement using a stabilizer simulator on Qiskit for the noiseless reference.
Note that although the qubits are not entangled to compare the results with the noiseless references, we can trivially use the entangled initial state by adding two-qubit gates in $U_{\mr{p}}$.

Figure~\ref{FIG4_Large_system_result_2column.pdf} shows the results of $\langle H \rangle$. 
The figures (a) and (c) represent the energy values for the 28- and 40-qubit models, respectively. 
For both models, the classically calculated values via the stabilizer simulator (blue line) fluctuate around the true Hartree--Fock energy $\langle H \rangle=\bra{{\rm HF}}H\ket{{\rm HF}}$ (red dashed line) as the number of shots increases. 
The mean values with standard deviations for the 28- and 40-qubit models using the results between  $10^6$ to $10^7$ shot counts are $-24.008\pm0.103$ and $-37.248\pm0.462$ Hartree, where Hartree--Fock energies are $-23.975$ and $-37.910$ Hartree, respectively.
Figures~\ref{FIG4_Large_system_result_2column.pdf}(b) and (d) show the absolute error from the Hartree--Fock energy for the 28- and 40-qubit models, respectively. In both models, as the number of shots increases, the errors tend to decrease. However, the error remains orders of Hartree, much larger than the chemical precision ($\approx 0.0016$ Hartree).
Also, the error in the 40-qubit model is larger than that in the 28-qubit model. 
The increase of locality in Pauli terms and the norm of a qubit Hamiltonian would be a cause of decreasing precision. It is noted that faster convergence could be achieved through more sophisticated qubit mappings or randomized methods~\cite{Bravyi2002-is, Hadfield2020-cr,Bertoni2022-kg,Zhao2021-vx} while this is out of scope in this study as the focus is on accelerating real-device implementation.

The real device results are shown by green lines in the figures; the mean values are $-20.091\pm0.054$ and $-31.113\pm0.803$ Hartree for the 28- and 40-qubit models, respectively.
Certainly, Figs.~\ref{FIG4_Large_system_result_2column.pdf}(a) and (b) show that there is a deviation from the true value. However, Figs.~\ref{FIG4_Large_system_result_2column.pdf}(c) and (d) indicate that the values converge around after $10^6$ shots. 
This convergence behavior becomes apparent in the regime of many sampling, which is in our case over $10^6$ sampling, highlighting an important insight enabled by the acceleration of the random algorithm in this study.
Note that the execution times for 28- and 40-qubit models are 182 and 262 hours, respectively. The slowdown in computation with increasing qubit count is considered to result from feedback operations, which should theoretically be executed in parallel but are instead executed sequentially. That is, the depth of feedback operations scales as $\order{n_\mr{s}}$ in the serial case and $\order{1}$ in the parallel case. According to recent updates~\cite{UnknownUnknown-sv}, this issue has been resolved in IBM devices, which means our proposal that the computational cost can be reduced depending on the backend.

Finally, we mention that we executed a readout error mitigation for these results to reduce the biases; see Fig.~\ref{FIG10_Error_mitigation_2column.pdf} in Appendix~\ref{sec: results for a readout error mitigation.}. 
The estimated values of energy via our method for 28 and 40 qubit models are $-20.111\pm0.056$ and $-31.475\pm1.118$ Hartree, 20 and 362 milli Hartree reduction from the original results, respectively. 
However, the large biases still remained, and thus the other types of errors such as the initialization, one-qubit gate, and dephasing errors, should be mitigated to obtain more accurate results~\cite{Zhao2024-by, Jnane2024-zv}.
It also suggests that parallel feedback operation is crucial for accurate computation.

\section{\label{sec: conclusions}Conclusions}

We proposed a dynamic-circuit-based method to efficiently execute randomized algorithms on real quantum devices. 
This method consists of generating the probability distribution in the amplitude of a quantum register state, recording the results in the classical register via measurement, and feeding them back to the quantum register to realize the randomized algorithm. 
We applied the proposed method to verify a one-qubit randomized Pauli measurement and found 14,000 times acceleration compared to the conventional method. 
The one-qubit randomized Pauli measurement can be utilized to realize the classical shadow method for observable estimation. 
We particularly computed Hartree--Fock energies for 28- and 40-qubit hydrogen chain models using the classical shadow on real devices. 
The method was successfully executed up to $10^7$ shots; this number of random circuit sampling is the largest compared with the previous studies~\cite{Elben2020-qt, Levy2024-de, Zhang2021-zf, Struchalin2021-bc, Huggins2022-ly, Korhonen2024-cp}.

Since randomized algorithms are utilized in a wide range of quantum algorithms, the applications of the proposed method are numerous. 
For example, potential applications in error mitigation include gate error mitigation techniques such as probabilistic error cancellation and readout error mitigation like twirled readout error extinction (TREX)~\cite{Temme2017-hf, Endo2018-nt, van-den-Berg2022-sr}. 
Also, the early fault-tolerant quantum computing approaches often leverage Monte Carlo sampling to reduce gate costs, e.g., in the problem of calculating expectation values for time-evolved states~\cite{Yang2021-ie}. 
Circuit cutting may also benefit from this acceleration, potentially enabling more two-qubit gates to be cut in practical scenarios. 

Methods of embedding a probability distribution into a circuit are becoming increasingly important for various applications~\cite{Nakaji2022-xe,Mitsuda2024-bt,Shirakawa2024-yi}. 
While conventional distribution embedding often assumes quantum variational algorithms, the proposed method may offer new insights because it allows classical distribution embeddings to a classically prepared shallow circuit~\cite{Shirakawa2024-yi,Ran2020-li,Rudolph2024-jo,Kanno2025-tt}.
As these methods are developing smoothly, we anticipate that dynamic-circuit-based approaches will accelerate existing quantum algorithms in various sense and open new pathways toward future quantum computing.
Furthermore, it will open up the development of quantum algorithms that leverage probability distributions, which are difficult to generate classically.

\textit{Note added}: during the review of this manuscript, AWS Braket Program Sets~\cite{Unknown2025-ks} was announced. The techniques for accelerating transpilation would also be important for reducing overall execution time.

\section{\label{sec:Data availability}Data availability}
The part of the codes and datasets in this study is available at \url{https://github.com/sk888ks/Random_algorithm_dynamic_circuit_open.git}.

\appendices

\begin{table}[h]
\centering
\caption{Values of $S$ for simulation time $t$ on the 28- and 40-qubit hydrogen chain models with $\epsilon = 0.01$.}
\label{tab: execution time}
\begin{tabular}{cll}
\hline
$t$ & 28 qubits & 40 qubits \\
\hline
$10^{0}$  & $3.6\times10^{6}$  & $1.6\times10^{7}$  \\
$10^{1}$  & $3.6\times10^{8}$  & $1.6\times10^{9}$  \\
$10^{2}$  & $3.6\times10^{10}$ & $1.6\times10^{11}$ \\
$10^{3}$  & $3.6\times10^{12}$ & $1.6\times10^{13}$ \\
\hline
\end{tabular}
\end{table}

\section{\label{sec: construction example of qDRIFT}Construction example of qDrift}

We describe the detailed procedure for implementing qDrift in Fig.~\ref{FIG2_Method_less1.5column.pdf}(c).  
In this procedure, $G_j$, $F_j$, $C_j$, and $n_{\mr{a}j}$ are independent of $j$, and are therefore denoted as $G_{\mr{q}}$, $F_{\mr{q}}$, $C_{\mr{q}}$, and $n_{\mr{aq}}$ respectively. The procedure is the following four steps;

(1) On a classical computer, the coefficients of the Hamiltonian are normalized and stored as a probability distribution. Specifically, the probability of index $i$ is $p_i = h_i / \lambda$, where $\lambda = \sum_i h_i$. The number of indices is $C_{\mr{q}}$.

(2) On a classical computer, a quantum circuit $G_{\mr{q}}$ is constructed such that the measurement probabilities of its components match the probability distribution obtained in step (1), i.e.,
$p_i \approx \abs*{\bra{\Vec{i}} G_{\mr{q}} \ket{0}^{\otimes n_{\mr{a}j}}}^2$, where we assumed that the index $i$ can be corresponded to the measured bitstring $\Vec{i}$, and $n_{\mr{aq}} = \lceil\log_2 C_{\mr{q}}\rceil$.  
Specific construction methods include synthesizing approximate circuits through optimization~\cite{Nakaji2022-xe} or implementing circuits via embedding the distribution into a matrix product state (MPS)~\cite{Han2018-bh}.  
The circuit depth depends on the construction method; for the MPS-based approach, it scales as $\mathcal{O}(\log n_{\mr{aq}})$~\cite{Malz2024-iv}.  

(3) On a quantum computer, the distribution is generated and stored in classical bits.  
The number of slots $S$ is given by $S=\lceil2\lambda^2 t^2/\epsilon \rceil$ in the original qDrift paper, where $t$ is a simulation time and $\epsilon$ is a desired precision. 
For each $j$, the distribution is generated and measured using $G_{\mr{q}}$, and the resulting bitstring is stored in $n_{cj}$ classical bits.

(4) On a quantum computer, after preparing the initial state, the results are fed back to the system qubits by $F_{\mr{q}}$.  
For each $j$, the quantum operation corresponding to $\exp(i \lambda t H_i /S)$ is applied according to the bitstring stored in the $n_{cj}$ classical bits.

We confirmed that the circuit size of step (2) is classically tractable for hydrogen chain models:  
the values of $C_{\mr{q}}$ are 27735 and 116597, and corresponding $n_{\mr{a}j}$ are 15 and 17 for the 28- and 40-qubit models, respectively.

The values of $S$ are shown in Table~\ref{tab: execution time} for reference.
While the original qDrift algorithm has a high gate cost, it can be significantly improved by applying Richardson extrapolation techniques~\cite{Watson2024-fj}.

\begin{figure*}[]
    \centering
    \includegraphics[width=0.7\textwidth]{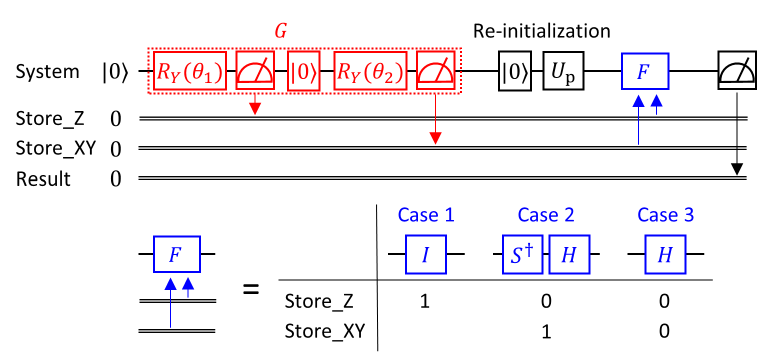}
    \caption{The naive implementation for the random Pauli measurement. The representations in this figure are the same as that in Fig.~\ref{FIG3_One_qubit_test_1column.pdf}. The cases in the lower panel are determined by the value of $\mr{Store\_Z}$ and $\mr{Store\_XY}$. } 
    \label{FIG5_One_qubit_test_original_1column.pdf}
\end{figure*}

\section{\label{sec: straightforward implementation for random Pauli measurement}Naive implementation for random Pauli measurement}
Figure~\ref{FIG5_One_qubit_test_original_1column.pdf} shows the naive implementation for random Pauli measurement. The number of slots $S$ is one, the number of cases $C_1$ is three, and the classical bit count $n_{\mr{a}1}$ is $n_{\mr{a}1} = \lceil \mr{log_2} C_1 \rceil = 2$. The lower panel shows the feedback operations for the cases.

\begin{figure}[]
    \centering
    \includegraphics[width=1\columnwidth]{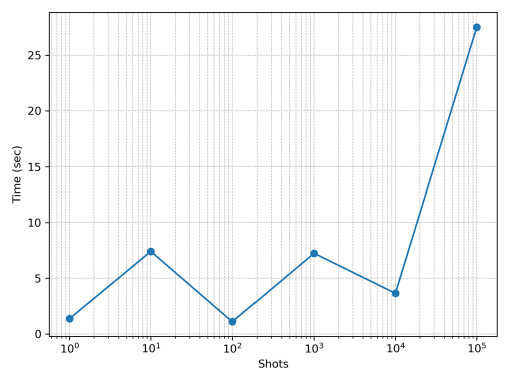}
    \caption{Execution times for shots. We mention that the execution was done by the latest Heron version of the $ibm\_kawasaki$ device, not the Eagle version in the main text.}
    \label{FIG6_Exectime_vs_shots_1column.pdf}
\end{figure}

\section{\label{sec: execution time analysis}Execution time analysis}

We executed jobs for the circuit in Fig.~\ref{FIG3_One_qubit_test_1column.pdf}(a) by changing the shot counts.
Figure~\ref{FIG6_Exectime_vs_shots_1column.pdf} shows the results of execution times for shots in the Heron version of the $ibm\_kawasaki$ device. Here we assume the execution time $T$ can be decomposed as $T = T_{c} + sT_{q}$, where $T_{c}$, $s$, and $T_{q}$ are compilation time, shots, and quantum device execution time, respectively~\cite{Karalekas2020-kd}. Since the fluctuation of execution time from $10^0$ to $10^4$ is not so large, we can regard $T_c >> T_q$. Therefore, the slowdown in the conventional approach is likely due to the compilation time as expected. Note that the time increases sharply at $10^5$ shots, which may be caused by congestion of the memory on the quantum control system~\cite{UnknownUnknown-sy}.

\begin{figure}[]
    \centering
    \includegraphics[width=1\columnwidth]{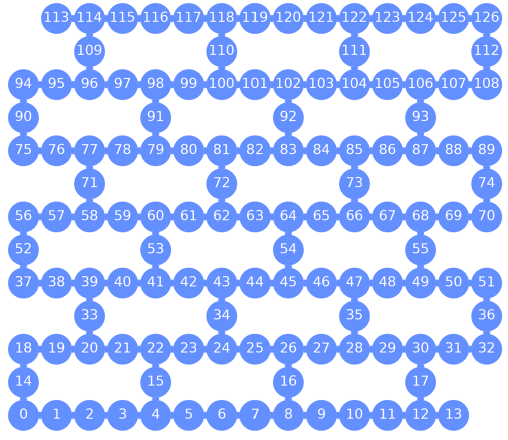}
    \caption{Device topology of the Eagle devices.}
    \label{FIG7_Device_topology_1column.pdf}
\end{figure}

\begin{figure*}[]
    \centering
    \includegraphics[width=1\textwidth]{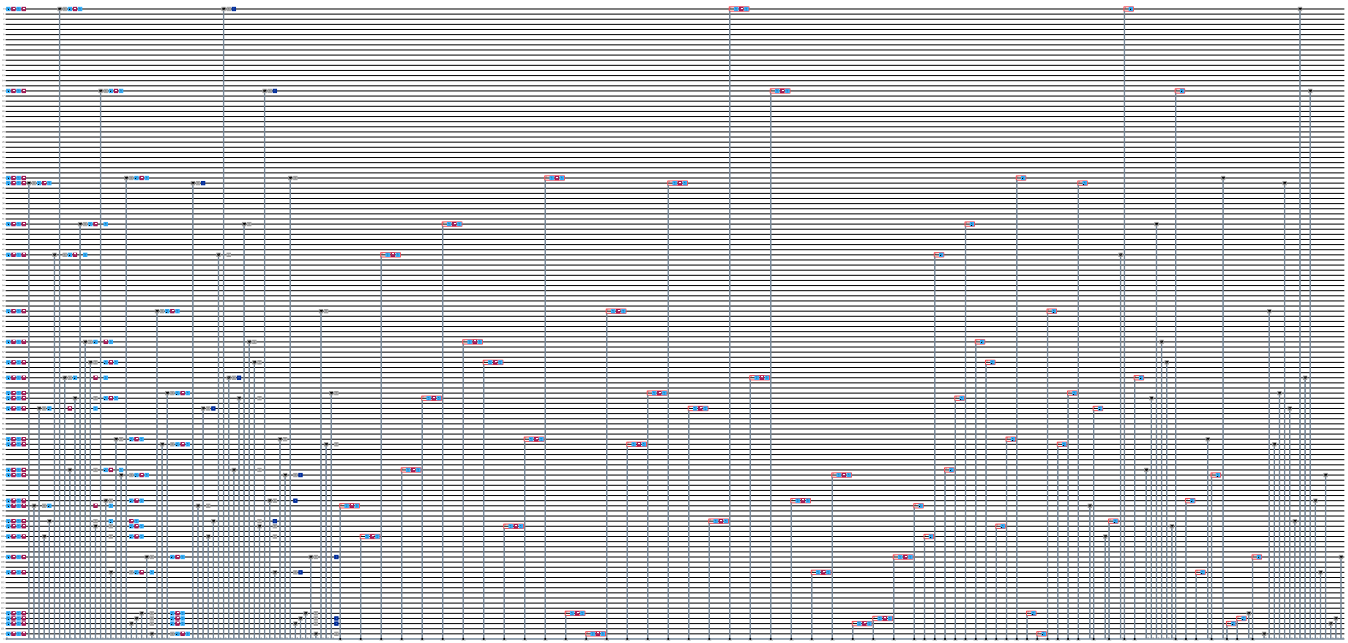}
    \caption{Circuit example in the 28 qubit model.}
    \label{FIG8_Circuit_28qubit.pdf}
\end{figure*}

\begin{figure*}[]
    \centering
    \includegraphics[width=1\textwidth]{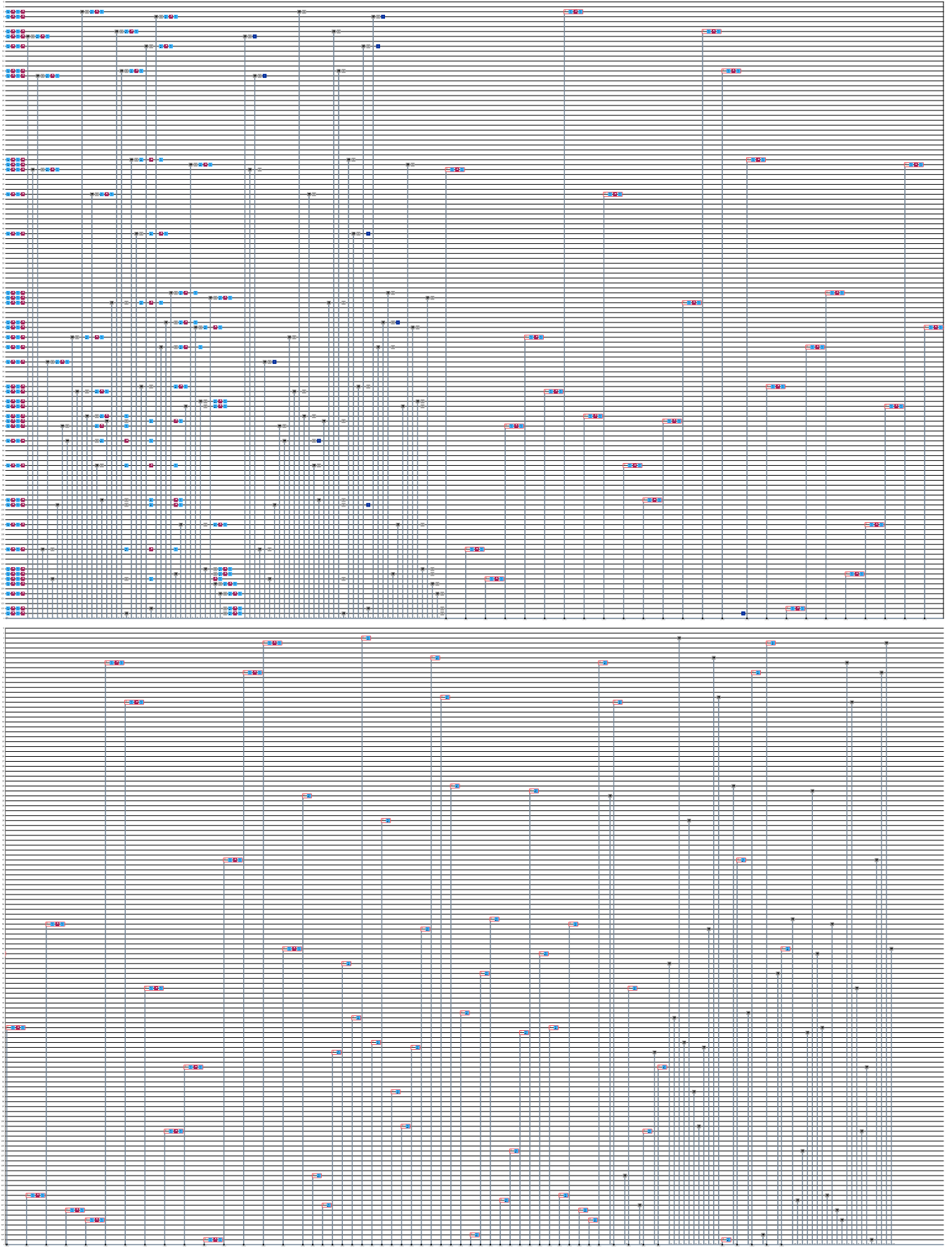}
    \caption{Circuit example in the 40 qubit model.}
    \label{FIG9_Circuit_40qubit.pdf}
\end{figure*}

\begin{figure*}[]
    \centering
    \includegraphics[width=1\textwidth]{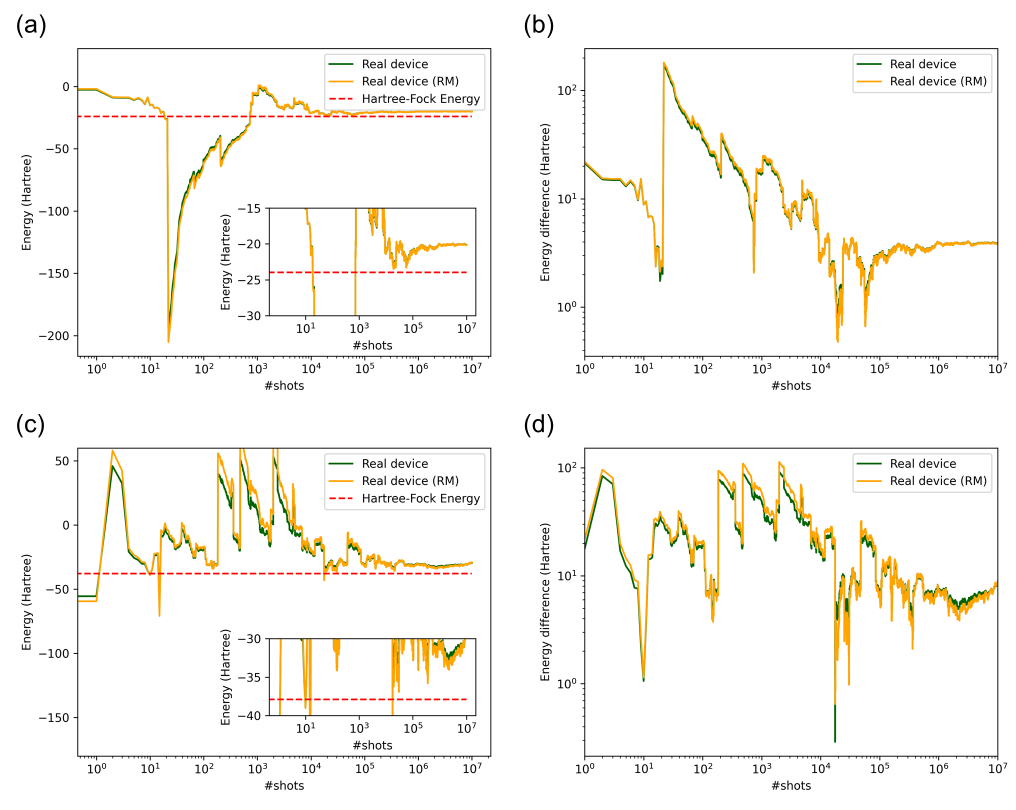}
    \caption{
    Results of expectation values for the hydrogen chain model with Hartree--Fock state including the readout error mitigation. 
    (a) and (b) [(c) and (d)] are the results for 28 [40] qubits. 
    (a) and (c) show the exact values of energy (Hartree-Fock energy) $\bra{{\rm HF}}H\ket{{\rm HF}}$ as well as the estimated values via the real devices with and without the readout error mitigation (RM). 
    (b) and (d) show the results of differences between the exact values and the estimated values in the log-log plot. 
    We adopt $ibm\_kawasaki$ and $ibm\_kyiv$ devices for the cases of 28 and 40 qubits demonstration, respectively.}
    \label{FIG10_Error_mitigation_2column.pdf}
\end{figure*}

\section{Circuit in large scale executions.\label{sec:large circuits}}
Figure~\ref{FIG7_Device_topology_1column.pdf} shows the device topology of the Eagle devices.
Figures~\ref{FIG8_Circuit_28qubit.pdf} and~\ref{FIG9_Circuit_40qubit.pdf} are circuit examples for 28 and 40-qubit models, respectively, where the qubit indices are shared with Fig.~\ref{FIG7_Device_topology_1column.pdf}.

\section{Results for a readout error mitigation.\label{sec: results for a readout error mitigation.}}

We tried readout error mitigation based on the readout error from calibration data.

The algorithm used in the present paper is described as Algorithm \ref{alg:shadow-mitigation}.
The notations here (specifically $Q, \alpha_Q, f(P,Q),$ and $\mu(P,\supp(Q))$) follow those of Ref.~\cite{Hadfield2020-cr}.
Let the readout error for a qubit $i$ be denoted as $e(i)$ and the corresponding factor is $m(i)=\frac{1}{1-2e(i)}$.
This is based on the idea that, assuming a symmetric error on a single qubit, the reduction in the expectation value is $1 - 2e$.
To restore the original expectation value, we multiply by its inverse.
For a subset $A\subseteq \set{1,2,\dots,n}$, $m(A)$ is defined as $m(A) = \prod_{i\in A} m(i)$

\begin{algorithm*}
\caption{Classical shadows with single qubit readout error mitigation}
\label{alg:shadow-mitigation}
    \For{\textbf{all} sample $s \gets \{1, 2, \dots, S\}$}{
            Prepare state $\rho$\;
            Random pick $P \gets \{X,Y,Z\}^{\otimes n}$\;
            \For{\textbf{all} qubit $i \gets \{1,2,\dots, n\}$}{
                    Measure qubit $i$ in $P_i$ basis providing eigenvalue measurement $\mu(P,i) \in \{\pm 1\}$\;
                }%\EndFor
            Estimate expectation value
            \[\nu^{(s)} = \sum_Q \alpha_Q m(\supp(Q)) f(P,Q) \mu(P,\supp(Q)) ;\]
        }%\EndFor
    \Return $\nu = \frac{1}{S} \sum_s \nu^{(s)}$\;
\end{algorithm*}

Figure~\ref{FIG10_Error_mitigation_2column.pdf} shows the results for the classical shadow before (green) and after (orange) the readout error mitigation.
The improvement by the error mitigation was small.
This could be due to differences in error rates between calibration and job execution, or the presence of significant correlated readout errors among multiple qubits.
It is possible that the error during execution was larger than the error during calibration.
To address this, it will be essential to measure the error rate during execution.
Furthermore, by symmetrizing the errors, it should be possible to more efficiently estimate their impact on the estimation value~\cite{van-den-Berg2022-sr}.
Our quantum random number generator could also be leveraged in this measurement symmetrization process.
This method takes account of the contribution of correlated errors through symmetrization.

\section*{\label{sec:Acknowledgments}ACKNOWLEDGMENT}
The part of calculations was performed on the Mitsubishi Chemical Corporation (MCC) high-performance computer (HPC) system “NAYUTA”, where “NAYUTA” is a nickname for MCC HPC and is not a product or service name of MCC.
We acknowledge the use of IBM Quantum services for experiments in this paper. The views expressed are those of the authors, and do not reflect the official policy or position of IBM or the IBM Quantum team.

\ifCLASSOPTIONcaptionsoff
  \newpage
\fi

\clearpage
\bibliographystyle{ieeetr}
\bibliography{RADC}

\begin{IEEEbiographynophoto}{Shu Kanno}
received a B.S. in Engineering from Tokyo University of Agriculture and Technology, Japan, in 2016, and a Ph.D. in Science from Tokyo Institute of Technology, Japan, in 2022.
Since 2022, he has been a researcher at Mitsubishi Chemical Corporation. The research interests include chemical applications of quantum computers.
\end{IEEEbiographynophoto}

\begin{IEEEbiographynophoto}{Ikko Hamamura}
He received his B.E., M.E., and Ph.D. degrees
in engineering from Kyoto University, Kyoto
Japan, in 2015, 2017, and 2020, respectively.
Since 2020, he has been a research scientist at IBM Quantum, IBM Research – Tokyo.
His research interests include quantum informition theory, foundation of quantum physics, and quantum computing.
After this work, he joined NVIDIA in 2024.
\end{IEEEbiographynophoto}

\begin{IEEEbiographynophoto}{Rudy Raymond}
received the B.Eng. degree in computer science and the M.Sc. and Doctoral degrees in informatics from Kyoto University, in 2001, 2003, and 2006, respectively. He has been a Researcher at IBM Research Tokyo in Japan since 2006, a Project Researcher at the Quantum Computing Center, Keio University, since 2018, and a part-time Lecturer at the University of Tokyo since 2019. He was appointed as a Consulting Professor at the Department of Computer Science, The University of Tokyo, in August 2022.
\end{IEEEbiographynophoto}

\begin{IEEEbiographynophoto}{Qi Gao}
received his B.S. in Life Science and Technology, and his M.S. and Ph.D. in Biological Resources and Informatics from the Tokyo Institute of Technology in 2005, 2007, and 2010, respectively. He is currently a Senior Chief Scientist at Mitsubishi Chemical Corporation and a Researcher at the Quantum Computing Center, Keio University. His work focuses on developing cutting-edge quantum methods to drive revolutionary advancements in computational chemistry and materials science.
\end{IEEEbiographynophoto}

\begin{IEEEbiographynophoto}{Naoki Yamamoto}
received the B.S. degree in engineering and the M.S. and Ph.D. degrees in information physics and computing from The University of Tokyo in 1999, 2001, and 2004, respectively. He is currently a Professor with the Department of Applied Physics and Physico-Informatics, Keio University, and the Chair of the Keio Quantum Computing Center. He was a Post-Doctoral Fellow with the California Institute of Technology from 2004 to 2007 and the Australian National University from 2007 to 2008. His research interests include quantum computation and control.
\end{IEEEbiographynophoto}

\end{document}